\documentstyle[twoside,fleqn,psfig,espcrc2]{article}
\newcommand{\be}{\begin{equation}}
\newcommand{\ee}{\end{equation}}
\newcommand{\bea}{\begin{eqnarray}}
\newcommand{\eea}{\end{eqnarray}}
\newcommand{\no}{\noindent}

\newlength{\figwidth}
\newlength{\figheight}
\title{A Study of Meson Correlators at Finite Temperature
\thanks{Poster by T. Umeda at LATTICE99.}}
\author{QCD-TARO Collaboration:
Ph.~de~Forcrand%
\address{ETH-Z\"urich, CH-8092 Z\"urich, Switzerland \\
$^{\rm b}$Dept. F\'{\i}sica Te\'orica, Universidad Aut\'onoma de Madrid,
      E-28049 Madrid, Spain \\
$^{\rm c}$Dept. of Appl. Phys., Fac. of Engineering,
           Fukui Univ., Fukui 910-8507, Japan \\
$^{\rm d}$Dept. of Physics, Tezukayama Univ.,Nara 631-8501, Japan \\
$^{\rm e}$Dept. of Physics, Hiroshima Univ.,
           Higashi-Hiroshima 739-8526, Japan \\
$^{\rm f}$Research Center for Nuclear Physics, Osaka Univ.,
           Ibaraki 567-0047, Japan \\
$^{\rm g}$Res. Inst. for Inform. Sc. and Education, Hiroshima Univ.,
           Higashi-Hiroshima  739-8521, Japan \\
$^{\rm h}$Institut f\"ur Theoretische Physik, Univ. Heidelberg
           D-69120 Heidelberg, Germany \\
$^{\rm i}$FEST, Schmeilweg 5, D-69118 Heidelberg, Germany \\
$^{\rm j}$Hiroshima University of Economics, Hiroshima 731-01, Japan },
M.~Garc{\'\i}a~P\'erez$^{\rm b}$,
T.~Hashimoto$^{\rm c}$,
S.~Hioki$^{\rm d}$,
R.~Katayama$^{\rm e}$,
H.~Matsufuru$^{\rm f}$,
O.~Miyamura$^{\rm e}$,
A.~Nakamura$^{\rm g}$,
I.-O.~Stamatescu$^{\rm h,i}$,
T. Takaishi$^{\rm j}$,
and  T.~Umeda$^{\rm e}$ }

\begin{document}

\begin{abstract}
We present results for mesonic propagators in temporal and 
spatial directions at $T$ below and above the deconfining transition 
in quenched QCD.
Anisotropic lattices are used to get enough information in the
temporal direction.
We use the Wilson fermion action for light quarks and Fermilab
action for heavy quarks.
\end{abstract}

\maketitle

\section{Introduction}
Hadronic properties are expected to change with increasing
temperature, especially near to and above the QCD phase transition.
Forthcoming heavy ion collision experiments will give the opportunity
to investigate them, and require theoretical
studies of signals of the phase transition and the QGP formation.
Lattice QCD simulations allow us to perform model independent non-perturbative
studies.

So far most calculations of hadronic masses at finite temperature have
handled only the screening mass \cite{screening} but not the pole mass, 
which has a fundamental importance.
In addition to the intrinsic ambiguity in defining pole masses at finite temperature 
the main difficulty comes from the shortness 
of the ``temporal'' (temperature) extension, $l_{\tau}=1/T$.

Here, we attack the problem by using optimised hadronic operators 
on anisotropic lattices. 
Our strategy is as follows: \\
(1) to use anisotropic lattices to obtain a good resolution 
in the temporal direction.\\
(2) to seek for a good operator which has a large overlap
with the state of interest at zero temperature.\\
(3) to observe temperature dependent effects by using this optimised 
operator on finite temperature lattices. 

In the following, we report results on 
light and heavy mesons.
Since part of the former have already been reported \cite{TARO99},
we briefly summarise the current results.
On the latter, 
we describe our computation and preliminary results. 
\section{Results on light mesons at finite temperature }
We use lattices of $12^3 \times N_{\tau}$
with $N_{\tau}=72, 20,16$ and $12$ at $\beta =5.68$, $\gamma_G=4$
,
in the quenched approximation
\footnote{The lattice mentioned in this report corresponds to
``Set-B'' data in \cite{TARO99}}.
Calibration using Wilson loops at $N_t=72$ gives the anisotropy
$\xi=a_{\sigma}/a_{\tau}=5.3(1)$ and $a_{\sigma}^{-1}=0.85(3)$ GeV
.
Our lattices correspond to temperatures  
$T \ \simeq 0, 0.93 T_c, 1.15 T_c$ and
$1.5T_c$ respectively.

For light quarks, anisotropic Wilson fermions are employed. 
After Coulomb gauge fixing, the PS meson is measured by the following 
operators;  
\begin{eqnarray}
G_{PS}(r,t)=
\sum_{\vec{m_1},\vec{m_2},\vec{n}}
w(\vec{m_1})w(\vec{m_2}) \nonumber \\
< Tr[M(\vec{m_1},0;\vec{n},t)M(\vec{m_2},0;\vec{n}+\vec{r},t)^
{\dagger}
> 
\label{opt}
\end{eqnarray}
where
$M(\vec{m},m_0;\vec{n},n_0)$ is a quark propagator and 
$w(\vec{m})$ is 
$exp(-\alpha|\vec{m}|^p)$, which we call  ``exp-source'' here.
After tuning the parameters $\alpha$ and $p$, 
the correlator at $T=0$ is dominated by the lowest state  
\cite{TARO99}.
Then the correlators at finite temperature are measured:
It turns out that the effective masses change significantly above $T_c$,
while $T \simeq 0.93T_c$ ($N_t=20$) reveals 
almost the same features as $T=0$.

\begin{figure}[tb]
\center{
\vspace{-5mm}
\leavevmode\psfig{file=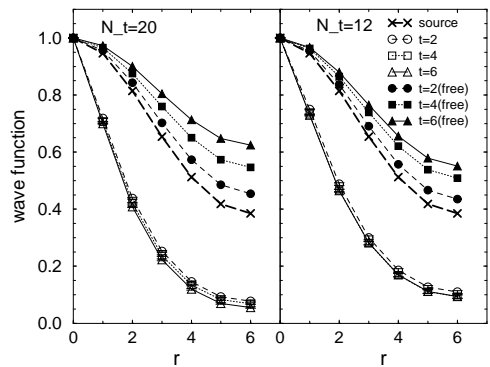,width=0.95\figwidth}}
\vspace{-11mm}
\caption{
Ps wave function ($\kappa_{\sigma}=0.086$, {\it exp-exp} source) 
normalised at $r=0$. 
Also plotted is the initial distribution of separations as given by the 
source.
$T \simeq 0.93T_c$ (left) and
$T \simeq 1.5T_c$ (right).}
\vspace{-3mm}
\label{fig:wave_func}
\end{figure}

We also examine the spatial structure of the correlator  
above $T_c$.
In Fig.~\ref{fig:wave_func} we compare the $Ps$
``wave functions" normalised at $x=0$,
$G_{Ps}(x,t)/G_{Ps}(0,t)$, at several $t$ for $T\! \simeq \!0.93T_c$
and $T\! \simeq \!1.5T_c$.
Also shown is the free quark case
($m_q a_{\sigma}=0.1$, $\gamma_F=\xi = 5.3$).
The {\it exp-exp} source appears somewhat too broad at $T \simeq 0.93T_c
$:
the quarks come close to each other while propagating in $t$.
An unexpected fact is that this is also the case at $T \simeq 1.5T_c$:
the spatial distribution shrinks and stabilises, indicating that even
at this high temperature there is tendency for quark and anti-quark
to stay together.
This is in clear contrast to the free quark case
which never shows such a behaviour regardless of the source.
The same feature holds for the other mesons at all quark masses.

\begin{figure}[tb]
\center{
\vspace{-5mm}
\leavevmode\psfig{file=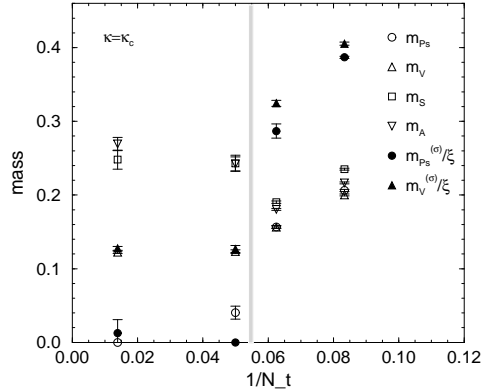,width=0.9\figwidth}}
\vspace{-11mm}
\caption{
Temperature dependence of the mass $m$ (open symbols)
and screening mass $m^{(\sigma)}$ (full symbols)
in units $a_{\tau}^{-1}$in the chiral limit.
The vertical gray lines indicate $T_c$. The data correspond to
$T \simeq 0,~0.93T_c,~1.15T_c$ and $1.5T_c$.  }
\vspace{-7mm}
\label{fig:mass}
\end{figure}

Fig.~\ref{fig:mass} shows the temperature dependence of ``masses''
which are extracted from the ``exp-exp'' correlators at
the three points $[N_t/2-2, N_t/2]$.
Though the values  above $T_c$ have large systematic errors, 
pole masses indicate
chiral restoration there.
We also show in the figure the screening masses: their difference with
the pole masses is manifest.
It is also noted that mesons degenerate approximately irrespective 
of their spin.
\section{Results on heavy quark systems at finite temperature}
Our target is mainly charmonium physics,
e.g. investigations of the mass shift near $T_c$ \cite{Has86}
and $J/\psi$ suppression in the plasma phase \cite{MS86}.
In order to treat heavy quarks with the currently available computer power,
we use the effective theory approach.
We adopt Fermilab action with $O(a)$ improvement \cite{EKM97}.
On the anisotropic lattice, the heavy quark operator is 
\footnote{This form differs from that of \cite{Kls99}}:
\begin{eqnarray}
  K(x,y) \!\!\! &=& \!\!\! \delta_{x,y}
       - \kappa_{\tau} \left\{ (1\!-\!\gamma_4)T_{+4}
                         +(1\!+\!\gamma_4)T_{-4} \right\}
 \nonumber \\
  & & \hspace{-20mm}
     - \kappa_{\sigma} \sum_{i} \left\{ (r-\gamma_i)T_{+i}
                    +(r+\gamma_i)T_{-i} \right\}
\nonumber \\ 
 & & \hspace{-20mm}
    -  r \left(  \kappa_t c_E g \sigma_{4i}F_{4i}(x)
               + \kappa_s c_B \frac{1}{2}
                  g \sigma_{ij}F_{ij}(x) \right) \delta_{x,y}.
 \label{eq:clover_aniso2} 
\end{eqnarray}
where,
$T_{\pm \mu} = U_{\pm \mu}(x) \delta_{x\pm\hat{\mu},y}$,
$\kappa_s = 1/2(m+\xi+3r)$, $\kappa_t = \xi \kappa_s$,
and $r=1/\xi$.
Mean-field improvement is done for $c_E$ and $c_B$.

Introducing the anisotropy results in changing the free quark
dispersion relation.
In eq.(\ref{eq:clover_aniso2}), one notices
that a larger anisotropy $\xi$ causes a smaller Wilson term.
Then the problem is how the contribution from 
the momentum region at the edge of the Brillouin zone becomes
larger.
In the heavy quarkonium, typical energy and momentum scales
exchanged in the meson are of order $mv^2$ and $mv$ respectively.
For the charmonium, $v^2 \sim 0.3$, and the typical energy scale
is $\sim 500$ MeV.
Let us consider our lattice: $a_{\sigma}^{-1}=0.85$
GeV and $\xi=5.3$, $m \sim 0.3$ roughly corresponds to the charm quark
mass.
From the dispersion relation, $E(p_z=a/\pi)-E(0) \sim 0.5$ GeV.
Though this result seems not sufficient, the effect is expected
to be small for the ground state.
With increasing lattice cutoff, this problem becomes milder.

We used the same configurations as for the light meson study.
The anisotropy $\xi$ is fixed by using the dispersion relation.
At present one set of parameters is available:
$\kappa_{\sigma}=0.0984$ and $\gamma_F=3.67$ which corresponds
to the vector meson mass $m_V\sim 3.10$ GeV.

\begin{figure}[htb]
\center{
\vspace{-0mm}
\leavevmode\psfig{file=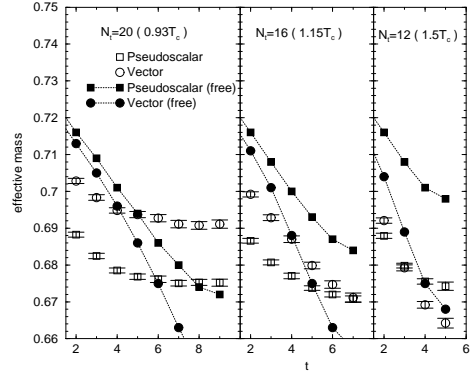,width=0.8\figwidth}}
\vspace{-8mm}
\caption{
Effective mass with $\kappa_{\sigma}=0.0984$ and 
$\gamma_F=3.67$. At $N_\tau = 72$ Pseudoscalar and Vector meson
mass are 0.6756(7) and 0.6890(8) respectively. Full symbols show
effective masses constructed by free quarks }
\vspace{-5mm}
\label{fig:heavy}
\end{figure}

Fig.\ref{fig:heavy} shows the effective masses of $Ps$
and $V$ mesons at finite temperature.
As for light mesons, on the $N_{\tau}=20$ lattice
the lowest vector and pseudoscalar
states keep almost the same masses as at $T=0$.
Above $T_c$ we observe, as in the light meson case, 
that the V meson effective mass goes below the Ps meson.
However, the Ps effective mass stays at almost the same position 
as at $T=0$ while the V effective mass decreases fast,
in contrast to the light meson case
where both of them increase with $T$.

\section{Summary and Outlook}

For a light quark system, the wave function indicates 
a strong correlation between quark and antiquark
even above $T_c$. 
The pole masses show chiral restoration above $T_c$ and a behaviour
quite different from that of the screening masses.
For heavy quark systems we use the Fermilab action 
on an anisotropic lattice. 
The effective mass of the vector meson drops sharply above $T_c$.
Further investigations with finer lattices are undergoing.

The calculations have been done 
Intel Paragon XP/S at INSAM, Hiroshima Univ.
This work is supported by the Grant-in-Aide for Scientific Research 
by Monbusho,Japan (No 10640272, No 11440080).


\begin{thebibliography}{9}

\bibitem{screening}C.DeTar and J.Kogut, {\it Phys. Rev.} {\bf D36} (1987) 2828;
C.DeTar, {\it Phys. Rev.} {\bf D36} (1988) 2328. 
\bibitem{TARO99}
 QCD-TARO: Ph. de Forcrand et al.,
 Nucl. Phys. B(Proc. Suppl.) {\bf 73} (1999) 420:
 hep-lat/9809173;
 QCD-TARO: Ph. de Forcrand et al.,
 hep-lat/9901017
\bibitem{Has86}
 T. Hashimoto et al., 
 Phys. Rev. Lett. 57 (1986) 2123-2126.
\bibitem{MS86}
 T. Matsui, H. Satz,
 Phys. Lett. B178 (1986) 416-422.
\bibitem{EKM97}
 A. El-Khadra, A.S. Kronfeld and P.B. Mackenzie, 
 Phys. Rev. D55 (1997) 3933-3957.
\bibitem{Kls99}
 T.R. Klassen,
 Nucl. Phys. B(Proc. Suppl.) {\bf 73} (1999) 918-920:
 hep-lat/9809174 
\end{thebibliography}
\end{document}